\documentclass[twocolumn,showpacs,superscriptaddress,floatfix]{revtex4}

\usepackage{latexsym}
\usepackage{amssymb}
\usepackage{graphicx}

\begin{document}
%%%%%%%%%%%%%%%%%%%%%%%%%%%%%%%%%%%%%%%%%%%%%%%%%%%%%%%%

\title{Structural transitions in scale-free networks}

\author{G\'abor Szab\'o}
\affiliation{Department of Theoretical Physics, Institute of Physics,
Budapest University of Technology, 8 Budafoki \'ut, H-1111 Hungary}
\affiliation{Laboratory of Physics, Helsinki University of Technology,
P.O.Box 1100, FIN-02015 HUT, Finland}

\author{Mikko Alava}
\affiliation{Laboratory of Physics, Helsinki University of Technology,
P.O.Box 1100, FIN-02015 HUT, Finland}
\affiliation{NORDITA, Blegdamsvej 17, DK-2100 Copenhagen, Denmark}

\author{J\'anos Kert\'esz}
\affiliation{Department of Theoretical Physics, Institute of Physics,
Budapest University of Technology, 8 Budafoki \'ut, H-1111 Hungary}
\affiliation{Laboratory of Computational Engineering, Helsinki
University of Technology, FIN-02015 HUT, Finland}

\date{\today}

\begin{abstract}
Real growing networks like the WWW or personal connection based
networks are characterized by a high degree of clustering, in addition
to the small-world property and the absence of a characteristic
scale. Appropriate modifications of the (Barab\'asi-Albert)
preferential attachment network growth capture all these aspects. We
present a scaling theory to describe the behavior of the generalized
models and the mean field rate equation for the problem. This is
solved for a specific case with the result $C(k) \sim 1/k$ for the
clustering of a node of degree $k$. Numerical results agree with such
a mean-field exponent which also reproduces the clustering of many
real networks.
\end{abstract}

\maketitle 

In diverse fields of scientific interest underlying network structures
can be recognized, which provide a unifying concept of investigation
\cite{general}. Examples range from biology (metabolic
networks~\cite{metabolic}, protein nets in the cell~\cite{protein})
through sociology (movie actor
relationships~\cite{emergence_of_scaling}, coauthor
networks~\cite{newman}, sexual nets~\cite{liljeros}) to informatics
(Internet~\cite{faloutsos}, WWW~\cite{WWW}). In all these examples it
is easy to identify the constituents of the problem with the nodes of
a graph and their relationships with directed or undirected
links. During the last few years a great deal of information has
accumulated about such structures. Three apparent features seem to
characterize them rather robustly: i) a high degree of clustering,
i.e., if nodes $A$ and $B$ are linked to node $C$ then there is a good
chance that $A$ and $B$ are also linked; ii) the ``Small World''
property, i.e., the expected number of links needed to reach from one
arbitrarily selected node another one is low; iii) the absence of a
characteristic scale, which often appears so that the distribution
$P(k)$ of the degrees $k$ of nodes follows a power law.

It has been noticed from the beginning that clustering in real
networks is an essential and an almost ubiquitous feature. It measures
the deviation from a structure with vanishing correlations, and it has
been used to describe the tendency of networks to form cliques or
tightly connected neighborhoods. As an organizing principle, this is
most obvious in social networks, where connections are usually created
by personal acquaintances, like in the scientific collaboration
network. Considerable clustering has also been found in networks of
more diverse nature. Prime examples are the WWW, metabolic and protein
interaction networks, the actor network, the power grid of the United
States, the semantic web of English
words~\cite{hierarchical_organization}, and the backbone of the
Internet on both the autonomous system and the router
level~\cite{pastor,goh}. The number of entries in this list is on the
rise as new disciplines are being taken under consideration and raw
data are made available. A comprehensive examination of a variety of
real networks clustering can be found in
Ref.~\cite{hierarchical_organization}. In real networks, as a
combination of the properties i) and iii), the clustering coefficient
as a function of the degree of the nodes often follows a power law:
$C(k) \propto k^{-\alpha}$. The value of $\alpha$ is in many networks
close to 1.

In 1998 Watts and Strogatz created an interesting family of models:
introducing a rather low portion of random links between arbitrarily
selected pairs of nodes in a regular lattice has the consequence that
property ii) gets fulfilled while clustering does not decrease
considerably, assuring i)~\cite{footnote,watts_strogatz}. However, the
distribution of the degrees of nodes shows a characteristic peak
instead of the required power law. Barab\'asi and Albert (BA) realized
that in the examples mentioned at the beginning an important aspect is
that the networks are created by growth. BA proposed preferential
attachment (PA) as a growth rule: the new nodes are linked to the old
ones with a probability proportional to their the actual
degree~\cite{emergence_of_scaling}. The structures obtained this way
are scale free and have the Small World property. In spite of
capturing important aspects of growing networks, the clustering tends
rapidly to a constant as a function of the degree $k$ and vanishes in
the thermodynamic limit.

Recently, attempts have been undertaken to modify the PA network
growth models so as to increase clustering. In these models a
mechanism, controlled by a new parameter, is introduced to take into
account the effect that ``friends of friends get friends''. Indeed, it
has been possible to create models which have all the three properties
i)-iii)~\cite{tunable_clustering,hierarchical_organization,growing_scale_free}.

The aim of this paper is to present a general framework for the
transition from a PA graph with zero clustering to still scale free
graphs with $C(k) \propto k^{-\alpha}$, and to give a corresponding
mean-field (MF) and rate-equation theory. As an example we will take
the Holme-Kim model~\cite{tunable_clustering} (a modified BA one) for
which the MF rate equations can be solved exactly, leading to $\alpha
= 1$. This is also shown to describe the simulations very well. At the
end, we discuss the assumptions one needs to make, and how these
reflect the behavior of clustering in general.

We start from the simplest undirected BA model: a new node $j$ with
$m$ links is added to the system at (discrete) time $t$. A link from
node $j$ to node $i$ is drawn with probability $k_i/\sum k_i$. It is
known that the the average clustering at node $i$ is independent of
the degree $k_i$~\cite{growing_scale_free}:
\begin{eqnarray}
C(k_i) = \frac{m-1}{8} \, \frac{(\log N)^2}{N},
\label{local_c_for_BA}
\end{eqnarray}
i.e., it is inversely proportional to the number $N$ of nodes (with a
logarithmic correction)~\cite{wrongexpr}. For the generalization of
the BA model with enhanced clustering, we have a parameter $p$
representing an imposed tendency to form triangles on the graph. It is
chosen such that at $p=0$ the original BA model is recovered.

We propose as a scaling ansatz to describe the clustering coefficient
$C$ as a function of the degree $k$, the number of nodes $N$ and the
parameter $p$:
\begin{eqnarray}
C(k, N, p) = N^{-1} f \left(\frac{k}{k^*(N,p)} \right),
\label{scaling_ansatz}
\end{eqnarray}
where $f(x)$ is a scaling function with $f(x) \to \rm const.$ for $x
\gg 1$ and $f(x) \to \rm x^{-\alpha}$ for $x \ll 1$ and the behavior
in Eq.~(\ref{local_c_for_BA}) is already taken into account by fixing
the exponent of the prefactor of $f$. The characteristic degree $k^*$
is a monotonously increasing function of $N$ for fixed $p$ and it
should decrease as $p$ goes to zero. A natural assumption is then:
\begin{eqnarray}
k^*(N,p) \sim N^\gamma p^\delta.
\label{kstar}
\end{eqnarray}
As for small $k$ the clustering $C$ in Eq.~(\ref{scaling_ansatz})
should go like $k^{-\alpha}$ and become independent of $N$, we have
$\gamma = 1/\alpha$. The exponent $\delta \alpha$ describes, how for
$N \to \infty$ the clustering $C$ approaches its limiting value zero
as $p$ goes to zero. If we accept that in most cases $\alpha=1$, there
is one exponent to be determined, say $\delta$. We now clarify the
origin of $\alpha =1$ and $\delta=1$ for the model employed.

For this purpose we write down the {\em rate equations} for the
clustering in a general form. We thus need to consider the average
rate of change
\begin{equation}
\partial_t n_i = 
R(k_i,p) \sum_{n \in \Omega} R(k_n, p),
\label{general_form}
\end{equation}
where $n_i $ is the number of connected neighbors of site $i$, and
$C_i = n_i / (k_i (k_i - 1) / 2)$. Here $R$ is the rate at which $i$
gets new links, and we allow, in analogy with the scaling ansatz
presented above, the rate to depend on both the degrees of the node in
question and the parameter $p$. This can be ``annealed'' or
``quenched'', depending on whether the parameter describes stochastic
rules (as in the example below) or a fixed property of each node $i$.
E.g., $R$ can simply follow from the preferential attachment rule.
$\Omega$ is the set of neighbors of node $i$ and the sum caters for
the probability that a new node linked to $i$ also links to one of the
neighbors of $i$. This increases $n_i$ and enhances clustering. In
order to make Eq.~(\ref{general_form}) more concrete we discuss the
triad formation model~\cite{tunable_clustering} as an example.

The complications in solving a rate equation like
Eq.~(\ref{general_form}) arise from the correlations that are embedded
between the degree of node $i$ and the properties of its
neighborhood. For the triad formation model, the rules consist of a BA
model extended by a triad formation step. Initially, the network
contains $m_0$ vertices and no edges, and in every time step a new
vertex is added with $m$ undirected edges. The $m$ edges are then
one-by-one subsequently linked to $m$ different nodes in the
network. One performs a preferential attachment step for the first
edge as defined in the BA model. With probability $p$, the second and
further edges edges are joined to a randomly chosen neighbor of the
node selected in the previous PA step. Alternatively, with probability
$1 - p$, a PA step is performed again.

\begin{figure}
\begin{center}
\includegraphics[height=3cm]{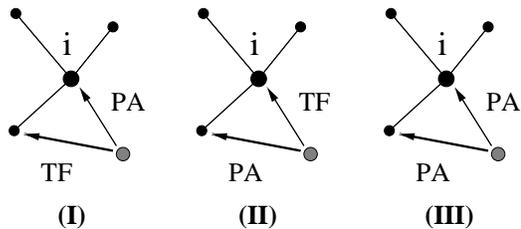}
\end{center}
\caption{Three different options to connect to node $i$ with $m \ge
2$. In (I), a PA step is performed first linking to $i$ and then a TF
step creates a link between neighbors of $i$. In (II), the same
happens, in a different order. (III) shows how two PA steps may
contribute to $n_i$. Bold edges increase $n_i$.}
\label{Figure1}
\end{figure}

In the limit when $p$ approaches zero, one recovers the original BA
model, and by setting $p$ to a value between $0$ and $1$ the average
clustering can be adjusted continuously and grows monotonously with an
increasing $p$. The microscopic mechanisms that increase $n_i$ are
illustrated in Figure~\ref{Figure1} and are (I): the new node connects
to node $i$ in a PA step, which is potentially followed by several TF
steps; (II): the new node connects to one of the neighbors of $i$ in a
PA step and then $i$ conversely gets linked to the new node in one of
the subsequent TF steps; (III): the new node connects to node $i$ in a
PA step and a neighbor of $i$ is also selected for connection to the
new node in another PA step.

Using these for $R(k_i, p)$, the rate equation for $n_i$ reads
\begin{eqnarray}
\partial_t n_i & = & m_{\mathit{PA}} \, \frac{k_i}{2mt} \, m_{\mathit{TF}} +
m_{\mathit{PA}} \sum_{n \in \Omega} \frac{k_n}{2mt} \,
\frac{1}{k_n} \, m_{\mathit{TF}} +
\nonumber\\
& + & \, m_{\mathit{PA}} \, \frac{k_i}{2mt} \,
(m_{\mathit{PA}} - 1) \sum_{n \in \Omega} \frac{k_n}{2mt}.
\label{n_i_rate_eq}
\end{eqnarray}

The first term in the sum gives the increase in $n_i$ by mechanism
(I). $m_{\mathit{PA}}$ is the number of PA steps attempted per each
new node (recall that per time-unit one new node is added).
$k_i/(2mt)$ is the preferential attachment probability to node $i$.
$m_{\mathit{TF}}$ is the expected number of triad formation steps that
take place on the average after a single PA step. Given this, we have
that $m_{\mathit{PA}} + m_{\mathit{PA}} m_{\mathit{TF}} = m$.

The second term describes mechanism (II); in this term, the sum goes
over all neighbors $\Omega$ of $i$, and their degrees are denoted by
$k_n$. $1/k_n$ comes from the fact that the neighboring node where a
TF step links is chosen uniformly from the neighbors. We exclude here
secondary triangle formation, that takes place if two TF steps from
the new node form a triangle with $i$ and one of $i$'s neighbors. This
becomes more relevant for large $p$'s. The term for (II) gives the
same expression as (I) after simplification.

The last term belongs to (III) and it is the only one that would
remain if we considered the simple BA model. It is the product of the
probabilities of linking to node $i$ and to one of the neighbors of
$i$, respectively, using only PA steps. The term contains the sum of
the degrees of neighboring nodes; this is $k_i$ times the average
degree of the neighbors. It has been shown that for uncorrelated
random BA networks $\langle k_n \rangle = \frac{\langle k \rangle}{4}
\log t = \frac{m}{2} \log t$~\cite{epidemic_structured_sf}. In this
model the numerical result follows the same scaling not only for $p
\ll 1$ but for $p$ general.

Finally, we get $n_i$ at the end of the network growth by integrating
both sides of Eq.~(\ref{n_i_rate_eq}). The integral for term (I) or
(II) is simply
\begin{eqnarray}
\int_1^N m_{\mathit{PA}} \, \frac{k_i}{2mt} \, m_{\mathit{TF}} \, dt =
\frac{m_{\mathit{PA}} m_{\mathit{TF}}}{m}
\int_1^N \frac{d k_i}{dt} \, dt =
\nonumber\\
= \frac{m_{\mathit{PA}} m_{\mathit{TF}}}{m} \, \left[ k_i(N) - m \right]
\approx \frac{m_{\mathit{PA}} m_{\mathit{TF}}}{m} \, k_i(N),
\label{I_II_integrated}
\end{eqnarray}
where we made use of the fact that $\partial_t k_i = k_i /
(2t)$~\cite{tunable_clustering}. From this, it also follows that
$k_i(t) = m (t/i)^{1/2}$ and thus integrating (III) gives
$[m_{\mathit{PA}} (m_{\mathit{PA}} - 1) / 16m] [(\log N)^2 / N]
k_i^2$. Combining this with Eq.~(\ref{I_II_integrated}) yields
\begin{eqnarray}
n_i = n_{i,0} + \frac{2 m_{\mathit{PA}} m_{\mathit{TF}}}{m} k_i +
 \frac{m_{\mathit{PA}} (m_{\mathit{PA}}-1)}{16m} \, 
\frac{(\log N)^2}{N} \, k_i^2.
\label{n_i_final}
\end{eqnarray}

The local clustering coefficient for node $i$ becomes
\begin{eqnarray}
C_i(k_i) = \frac{n_i}{k_i(k_i - 1) / 2} 
\approx \frac{4 m_{\mathit{TF}}}{k_i} +
\frac{m-1}{8} \, \frac{(\log N)^2}{N},
\label{local_c}
\end{eqnarray}
after neglecting $n_{i,0}$ and approximating $m_{\mathit{PA}}$ by $m$,
which is reasonable when the triad formation probability is small. It
is not surprising that the constant offset in the expression of $C_i$
is for $p\rightarrow 0$ exactly the constant clustering coefficient of
pure BA graphs. The first term, more importantly, can be attributed to
the triad formation induced clustering, and shows the $1/k$ behavior
typical of many real networks and other
models~\cite{hierarchical_organization,growing_scale_free,eckmann}.
$C_i$ is composed of a power law and a constant, so perfect power-law
behavior follows only when the former one dominates. In the opposite
case an effective exponent will be less than $1$. Furthermore, since
$n_{i,0}$ has been neglected, Eq.~(\ref{local_c}) and the inverse
proportionality apply to nodes with $k_i$ large enough, only.

For further progress $m_{\mathit{TF}}$, the expected number of links
created in the several possible TF steps after a PA step for a
particular node, needs to be approximated. Take $m-1$ edges to be
available for successive TF steps (this is an upper limit) and assume
node $i$ is not saturated yet as far as the connections to the
neighbors are concerned. This gives $m_{\mathit{TF}} =
\sum_{z=1}^{m-2} z p^z (1-p) + (m-1) p^{m-1} \approx p$ for $p$ small.

\begin{figure}
\begin{center}
\includegraphics[width=0.8\linewidth]{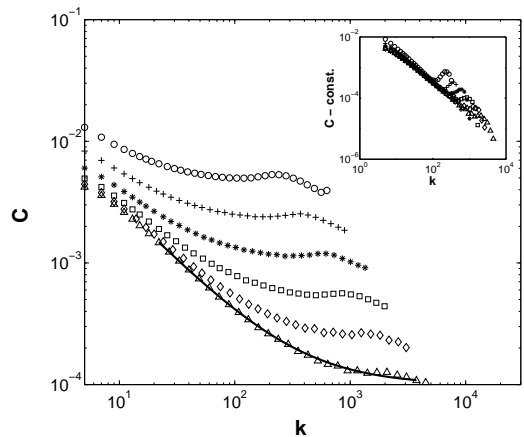}
\end{center}
\caption{Clustering coefficient as a function of the node degree for
$m=5$ and different sizes ($10^4$ for $\circ$, 25119 for $+$, 63096
for $*$, 158489 for $\Box$, 398107 for $\Diamond$, and $10^6$ for
$\triangle$). The triad formation probability is uniformly
$p=0.01$. The bold line is a least-square fit to the largest system
and gives $C \approx 0.028 \, k^{-0.97} + 9.9 \cdot 10^{-5}$, where
the prediction is that $C \approx 0.02 \, k^{-1} + 9.5 \cdot
10^{-5}$. The inset shows the data collapse of the power-law part of
$C(k)$.}
\label{Figure2}
\end{figure}

The fact that the local clustering coefficient contains a constant
term means that there is a crossover at a certain $k^*$. At this
point, a power law turns over to a constant clustering coefficient.
$k^*$ can be estimated by taking the two terms in Eq.~(\ref{local_c})
to be equal:
\begin{eqnarray}
%\frac{4 m_{\mathit{TF}}}{k^*} & = & \frac{m-1}{8} \, \frac{(\log N)^2}{N}
%\nonumber\\
k^* & \approx & \frac{32}{m (\log N)^2} \, pN.
\label{crossover_point}
\end{eqnarray}
Thus we can conclude that the exponents of Eq.~(\ref{kstar}) are
$\gamma=1/\alpha=1$ and $\delta=1$ for the triad formation model, and
from above, $\alpha=1$. E.g., in the case of Figure~\ref{Figure2},
taking $N = 10^6$ yields $k^* \approx 400$.

\begin{figure}
\begin{center}
\includegraphics[width=0.8\linewidth]{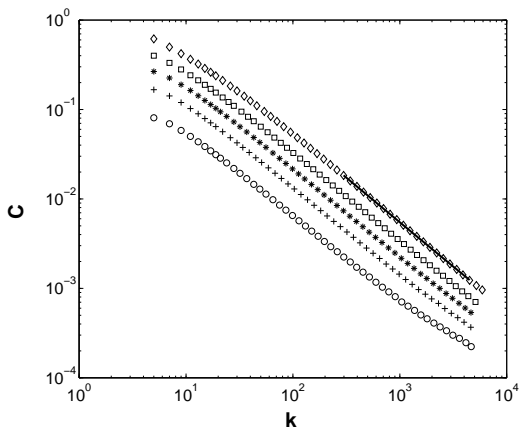}
\end{center}
\caption{Clustering coefficient for networks of $10^6$ nodes and
$m=5$; the triad formation probability is $p=0.2$, $0.4$, $0.6$,
$0.8$, and $1$, for $\circ$, $+$, $*$, $\Box$, and $\Diamond$,
respectively. The fit of Eq.~(\ref{local_c}) for $p=1$ gives $C
\approx 5.28 \, k^{-1}$, whereas the relation is expected to be $C
\approx 3.2 \, k^{-1} + 9.5 \cdot 10^{-5}$.}
\label{Figure3}
\end{figure}

Simulations of the model consistently confirm the analytical results
obtained from the rate equation. In Figure~\ref{Figure2} networks of
different sizes are shown to undergo such a transition to constant
clustering by tuning $p$ so that $k^*$ is smaller than the maximum
degree in the networks. A similar phenomenon to the transition
described above can be observed in the case of the actor network of
the IMDB database~\cite{hierarchical_organization}, where the tail of
a decreasing power law becomes constant, although large fluctuations
naturally affect this part of the statistics. Figure~\ref{Figure3}
shows networks well below the transition and thus almost only the
power-law part is conceivable.

It is not unusual in the physics of scale free networks that
mean-field approaches work well~\cite{general}. This fact is related
to the strongly hierarchical nature of the networks grown by
preferential attachment and our study demonstrates that this situation
remains unaltered even when considering a mechanism which enhances
clustering. The agreement between the $1/k^\alpha$ dependence with
$\alpha=1$ obtained in Eq.~(\ref{local_c}) and that found in real
networks indicates that the same ``mean-field'' mechanisms of
clustering are operative. For PA growth with enhanced clustering the
simplest interpretation is that for each new link a node $i$ gains
from a new node introduced to the network, its neigbors (``friends'')
have also a constant probability to be linked to the same new one.

It is interesting to ask how robust the mean-field exponent is and
what are the limits of the above approach, especially in the light of
the recently discovered networks with $\alpha \neq 1$~\cite{CCR}. The
rate equations allow to discuss the ways how exponents like such can
emerge. Eq.~(\ref{general_form}) implies that the clustering is
crucially dependent on the properties of the nodes in the
neighborhood, $\Omega$. If, say, correlations from ``assortative'' or
``disassortative'' mixing arise between $k_i$ and the average degree
$\langle k_n \rangle$ ($n \in \Omega$)~\cite{assortative_mixing}, this
may either enhance ($\alpha<1$) or inhibit ($\alpha>1$) clustering
from the mean-field result. On the level of models, one can envision
changing the $k$- and $p$-dependence of the rates. The second
possibility is fluctuation effects that limit the validity of the
rate-equation theory. It would seem to be of interest to explore both
these issues.

In conclusion, we have formulated a scaling assumption and a
mean-field theory of the clustering of scale-free networks. A specific
example, the triad formation model has been solved and comparisons to
the simulations indicate both good agreement and yield the MF-value of
the exponent $\alpha$. This approach should be amenable to many of the
models in the literature, and help in understanding the origins of the
statistical properties of clustering, also beyond the $C(k)$-function,
both in models and in the many real-life examples of networks. We have
here considered only growing networks, but obviously the rate
equations can be written down also in the case the structural dynamics
allows for deleting edges, as well \cite{CCR2,delete}.

Acknowledgements: JK and GS wish to thank for the warm hospitality at
HUT and for partial support by OTKA T029985. This work has been
supported by the Academy of Finland's Centre of Excellence Programme
and by the Centre for International Mobility (CIMO).

%%%%%%%%%%%%%%%%%%%%%%%%%%%%%


\begin{thebibliography}{99}
\bibitem{footnote} Of course, the regular lattice has to have high
clustering, e.g, by higher neighbor connections.
\bibitem{wrongexpr} Note that there is a factor of $m-1$, in contrast
to the original result of $m$ in \protect{\cite{growing_scale_free}}.
\end{thebibliography}
\end{document}